\documentclass[conference]{IEEEtran}
\IEEEoverridecommandlockouts
\usepackage{cite}
\usepackage{amsmath,amssymb,amsfonts}
\usepackage{graphicx}
\graphicspath{{./}{../}}
\usepackage{textcomp}
\usepackage{xcolor}
\usepackage{booktabs}
\usepackage{url}
\usepackage{stfloats}

\usepackage{eso-pic}
\begin{document}
\AddToShipoutPictureBG*{%
  \AtPageUpperLeft{\put(0,-28){\makebox[\paperwidth]{\small\itshape Accepted to the 2026 IEEE 15th International Conference on Consumer Electronics -- Berlin (ICCE-Berlin).}}}%
  \AtPageLowerLeft{\put(0,18){\makebox[\paperwidth]{\parbox[b]{\textwidth}{\scriptsize\centering \copyright~2026 IEEE. Personal use of this material is permitted. Permission from IEEE must be obtained for all other uses, in any current or future media, including reprinting/republishing this material for advertising or promotional purposes, creating new collective works, for resale or redistribution to servers or lists, or reuse of any copyrighted component of this work in other works.}}}}}

\title{Capacity over TFLOPS: Memory-Centric Deployment\\
of Large Clinical Language Models on Consumer CPUs
\thanks{Code, synthetic-data generator, and reproduction scripts:
\protect\url{https://github.com/balkeumlabs/LLM-on-CPU}}}

\author{\IEEEauthorblockN{Eli Yune}
\IEEEauthorblockA{\textit{Balkeum Labs} \\
Seongnam, South Korea \\
eli.yune@balkeumlabs.ai}
\and
\IEEEauthorblockN{Umer Majeed}
\IEEEauthorblockA{\textit{Balkeum Labs} \\
Seongnam, South Korea \\
umer.majeed@balkeumlabs.ai}
\and
\IEEEauthorblockN{Mahesh Balan}
\IEEEauthorblockA{\textit{Claremont Graduate University} \\
Claremont, CA, USA \\
balan.mahesh@cgu.edu}
}

\maketitle

\begin{abstract}
The dominant barrier to running large clinical language models at the point of
care is not arithmetic throughput but graphics memory: consumer GPUs cap at
4--16~GB, while the working set of a 27-billion-parameter model in 4-bit
precision is 17--25~GB. We argue that the edge node should be treated as
capacity-enabled (system RAM) rather than compute-limited (GPU VRAM), and we
test this on commodity hardware. Using an INT4 build of a 27B open model as an
architecture-faithful stand-in for a 27B medical model, we measure the
memory footprint, inference latency, energy, and federated communication cost
of a fully CPU-based pipeline driven by synthetic Fast Healthcare
Interoperability Resources (FHIR) prompts. On two independent consumer
machines---a desktop-class machine with 67~GB of RAM and a 6~GB GPU, and a
mobile workstation with 34~GB of RAM and a 4~GB GPU---the model loads entirely
into system RAM (37\% and 60\% utilization, respectively) while both GPUs are
out-of-memory before the first step. Inference runs at overnight-batch speed
(0.6--2.4 tokens/s for the 27B model), and federated fine-tuning would transmit
only a 22.6~MB low-rank adapter, a 99.96\% reduction over full weights. The
capacity result is independent of CPU microarchitecture, RAM size, operating
system, and GPU. We further identify an open, CPU- and NPU-optimized inference
runtime as the natural production path for this approach, indicating a
privacy-preserving and broadly deployable route for on-premises clinical AI.
All data is synthetic; this is a research artifact and not a medical device.
\end{abstract}

\begin{IEEEkeywords}
edge AI, large language models, CPU inference, quantization, low-rank
adaptation, federated learning, clinical natural language processing, memory
hierarchy
\end{IEEEkeywords}

\section{Introduction}

Deploying large language models (LLMs) inside a hospital or clinic is
attractive for privacy: patient records governed by HIPAA and GDPR never leave
the premises, and no protected health information is sent to a cloud provider.
The practical obstacle is hardware. The community reflex is to reach for a GPU,
but the relevant constraint for a high-parameter model is not floating-point
throughput---it is \emph{memory capacity}. A consumer GPU offers 4--16~GB of
VRAM, whereas the resident working set of a 27B model, even compressed to 4-bit
weights, is 17--25~GB once the key--value cache and runtime buffers are
included. The model simply does not fit, and the GPU reports an out-of-memory
(OOM) error before a single token is produced.

System RAM, by contrast, is cheap and abundant: 32--64~GB is unremarkable on a
commodity desktop or workstation. We therefore reframe the edge node as
\emph{capacity-enabled} rather than \emph{compute-limited}, and ask a direct
question: can a consumer CPU host a 27B clinical LLM in system RAM, and what is
the cost in latency, energy, and---for federated learning---network bandwidth?

This paper makes three contributions. (1) We show, on two independent commodity
machines, that an INT4 27B model loads and serves entirely from system RAM
while the on-board GPU cannot hold it; the result is robust to CPU microarchitecture, RAM
size, operating system, and GPU class. (2) We characterize the honest cost of
the CPU-only path---inference latency, throughput, and an energy proxy---using a
synthetic FHIR clinical workload. (3) We quantify the communication efficiency
of sharing only low-rank adapters in a federated setting. We are deliberate
about the boundary between measured and modeled quantities, and we make no
claim of any hardware-specific accelerator.

\section{Background and Related Work}

\textbf{Quantization.} Post-training quantization compresses LLM weights to 4
or 8 bits with modest quality loss. GPTQ~\cite{gptq} uses second-order
information for accurate 4-bit weights; activation-aware methods such as
AWQ~\cite{awq} protect salient channels; and SmoothQuant~\cite{smoothquant}
migrates activation outliers to enable 8-bit weight-and-activation inference.
These techniques are now routine. We use a 4-bit (Q4\_K\_M) GGUF build, which
reduces a 27B model from $\sim$54~GB (BF16) to $\sim$16.7~GB on disk.

\textbf{Parameter-efficient fine-tuning.} Low-Rank Adaptation
(LoRA)~\cite{lora} freezes the base weights and learns small low-rank update
matrices; QLoRA~\cite{qlora} demonstrated fine-tuning of quantized models. Only
the adapters need to be exchanged or stored, which is central to our
communication-efficiency analysis.

\textbf{Memory-efficient and offloaded inference.} Serving large models is
increasingly memory- rather than compute-bound. PagedAttention in
vLLM~\cite{vllm} manages the key--value cache in non-contiguous pages to curb
fragmentation, and FlexGen~\cite{flexgen} shows that aggregating CPU RAM (and
disk) lets a single commodity machine run models far larger than its GPU. Our
work takes this premise to its conclusion: drop the GPU entirely and treat
system RAM as the primary model store.

\textbf{Federated learning.} Federated averaging~\cite{fedavg} trains a shared
model across decentralized data without centralizing it---a natural fit for
clinical records. Secure aggregation~\cite{secagg} lets a server combine updates
without inspecting any single one, and federated instruction tuning of
LLMs~\cite{fedit} has been demonstrated by exchanging only adapter weights. Our
prior work, the FLAI Protocol~\cite{flai}, contributes a decentralized variant
with on-chain rewards and sPEG-based contribution scoring; the present study
supplies the missing local-compute substrate, showing that each participant can
hold the heavy base model in commodity RAM and train an adapter on-site.

\textbf{CPU and edge inference runtimes.} The \texttt{llama.cpp}
runtime~\cite{llamacpp} executes quantized models efficiently on CPUs across
operating systems, using system RAM for weights and cache. For production edge
deployment, the OpenVINO toolkit~\cite{openvino} provides a mature,
INT4/INT8-optimized inference runtime that targets CPUs, integrated GPUs, and
on-package neural processing units (NPUs) through a single API, with low-power
profiles well suited to bedside devices. We build our measurement harness on
\texttt{llama.cpp} because it is vendor-neutral and requires no GPU, and we
identify OpenVINO as the optimized runtime for hardware-accelerated deployment
(Sec.~\ref{sec:concl}).

\textbf{Clinical models and data.} Large models encode substantial clinical
knowledge: Med-PaLM~\cite{medpalm} approached expert performance on medical
question answering, and open efforts such as MEDITRON~\cite{meditron} and the
Gemma-derived MedGemma family~\cite{medgemma} bring this to deployable sizes;
the base Gemma-2 architecture~\cite{gemma2} is public. Standardized records
follow the HL7 FHIR specification~\cite{fhir}, and de-identified corpora such as
MIMIC-IV~\cite{mimic} are credentialed. To avoid any access or privacy concern,
we generate synthetic FHIR-shaped data and use an open Gemma-2 model as an
architecture-faithful stand-in for MedGemma; the memory footprint is determined
by parameter count and quantization, so it is representative.

\section{Methodology}

\subsection{Hardware}
We evaluate two independent consumer machines (Table~\ref{tab:hw}), procured and
operated separately, to test whether the capacity result is configuration
specific. The machines differ in CPU microarchitecture and generation, RAM size,
operating system, and GPU; one is a desktop-class platform and the other a mobile
workstation. We report the exact parts for reproducibility, but the study is
deliberately \emph{not} a vendor comparison---both are commodity CPUs, and the
question is whether such a CPU can do what the GPU cannot.

\begin{table}[t]
\caption{Test machines}
\label{tab:hw}
\centering
\begin{tabular}{@{}lll@{}}
\toprule
 & \textbf{Machine A} & \textbf{Machine B} \\
\midrule
Class    & Desktop & Mobile workstation \\
CPU      & AMD Ryzen 5 5500GT & Intel Xeon E-2176M \\
Cores    & 6C / 12T & 6C / 12T \\
RAM      & 67.2 GB & 34.1 GB \\
GPU      & 6 GB & 4 GB \\
OS       & Ubuntu 24.04 & Windows 11 \\
\bottomrule
\end{tabular}
\end{table}

\subsection{Software and models}
Inference uses \texttt{llama.cpp} (CPU backend) with INT4 (Q4\_K\_M) GGUF
weights. The capacity tier is Gemma-2-27B-it (16.65~GB on disk); a 2B edge tier
(Gemma-2-2B-it) is used for a like-for-like cross-machine comparison, since the
27B tier exceeds the smaller machine's comfortable RAM headroom only modestly
and the 2B tier runs on any laptop. No GPU offload and no
hardware-specific acceleration library is used on either machine; this gives a
conservative, portable baseline. An optimized production deployment would
instead use an INT4-aware CPU/NPU runtime such as OpenVINO~\cite{openvino}
(Sec.~\ref{sec:concl}); we keep that orthogonal to the capacity measurement here.

\subsection{Synthetic clinical workload}
We procedurally generate FHIR-shaped records---vital signs, laboratory panels,
problem lists, and medications---for 300 synthetic patients, partitioned into
five non-independent-and-identically-distributed (non-IID) specialty shards
(cardiology, pulmonology, nephrology, endocrinology, general) whose laboratory
abnormalities and diagnoses are deliberately skewed per specialty. From these we
derive 900 instruction prompts in three tasks (summarize vitals, flag abnormal
labs, triage acuity). Generation is seeded and reproducible; no real patient
data is used.

\subsection{Experiments and measurement honesty}
Four experiments are run by a single self-adapting script on each host:
\textbf{E1} (capacity) loads the model and records the resident set size and the
GPU verdict; \textbf{E2} (throughput) measures time-to-first-token (TTFT) and
tokens/s over the prompt set; \textbf{E3} (energy) estimates tokens-per-joule;
and \textbf{E4} (communication) computes the LoRA adapter size analytically.

We separate measured from modeled quantities. The GPU OOM verdict is
\emph{analytical}: the weight file (16.65~GB) strictly exceeds available VRAM
(6.4 and 4.3~GB), which guarantees failure before the first step; we do not
force a crash. Power is \emph{estimate-grade}: hardware energy counters (RAPL)
are root-only on these hosts, so E3 reports a labeled thermal-design-power
estimate. Throughput (E2) reflects the host CPU's class and memory bandwidth and
is reported as a vendor-neutral baseline, not an accelerator result.
Federated training itself is not executed; E4 is derived from the model
architecture.

\section{A Capacity-Enabled Federated Architecture}

\begin{figure*}[tb]
\centerline{\includegraphics[width=0.82\textwidth]{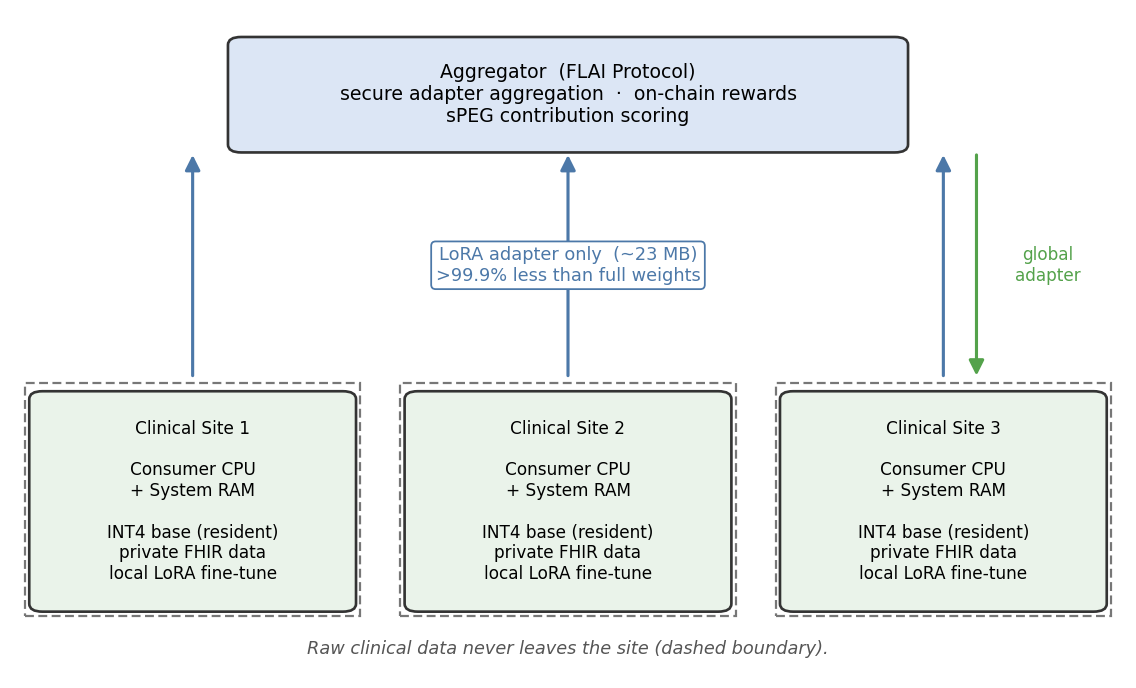}}
\caption{Capacity-enabled federated deployment. Each site holds the INT4 base
model in system RAM and fine-tunes a local adapter on private FHIR data; only
the small adapter is shared with the FLAI aggregator. Raw data never leaves the
site.}
\label{fig:arch}
\end{figure*}

The capacity result is not only a single-node curiosity; it is the enabling
condition for a privacy-preserving federated deployment (Fig.~\ref{fig:arch}).
Because each clinical site can hold the full quantized base model resident in
system RAM, fine-tuning can occur \emph{where the data already is}. Each site
trains a low-rank adapter on its local FHIR records; raw data never crosses the
site boundary. Only the adapter---22.6~MB at rank~16, versus tens of gigabytes
of base weights---is transmitted to an aggregator.

This is precisely the regime our FLAI Protocol~\cite{flai} was designed for: it
performs decentralized adapter aggregation with on-chain rewards and sPEG-based
contribution scoring, so that participating sites are incentivized and credited
in proportion to the value of their updates. The two lines of work are
complementary. FLAI specifies \emph{how} updates are aggregated, rewarded, and
trusted; the present work establishes \emph{that} the underlying per-site
computation---hosting and adapting a large clinical model---is feasible on
hardware a clinic already owns, with no GPU and no data egress. Combined, they
describe an end-to-end path from commodity edge hardware to an incentivized,
auditable, privacy-preserving clinical-intelligence network. Secure
aggregation~\cite{secagg} can further ensure the aggregator never inspects an
individual site's adapter.

\section{Results}

\begin{figure*}[tb]
\centerline{\includegraphics[width=0.92\textwidth]{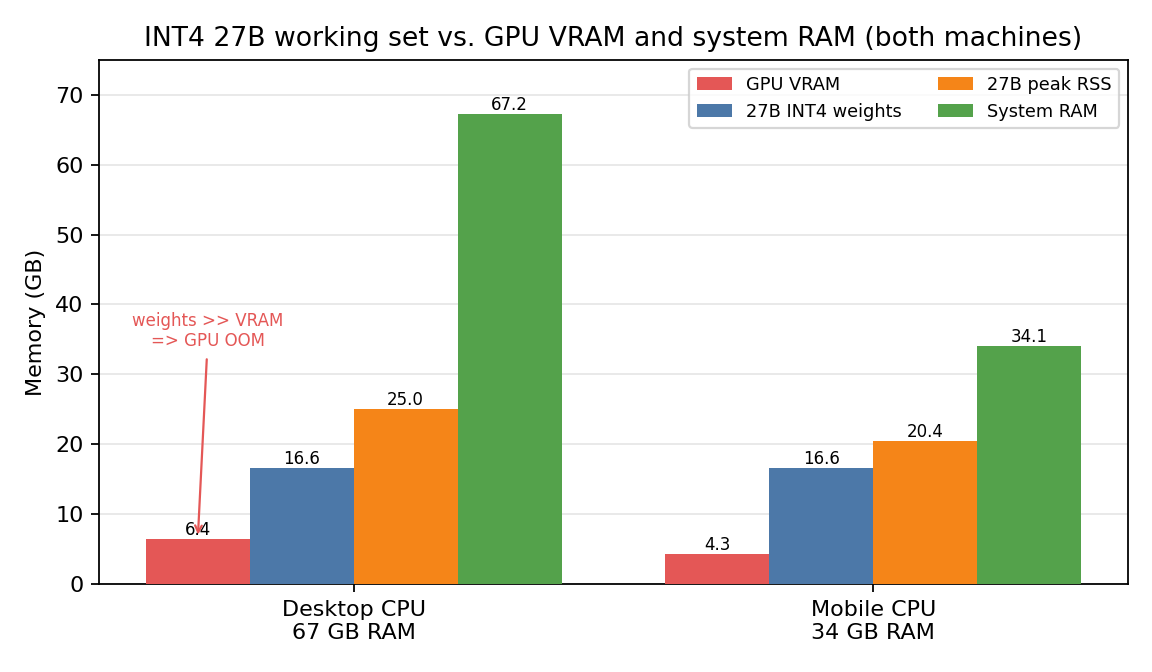}}
\caption{The 27B INT4 working set (weights and peak resident memory) far exceeds
each machine's GPU VRAM---guaranteeing an out-of-memory failure---yet fits
comfortably within system RAM on both machines.}
\label{fig:cap}
\end{figure*}

\subsection{E1: Capacity over TFLOPS}
The headline result is summarized in Fig.~\ref{fig:cap} and the top of
Table~\ref{tab:main}. On Machine A the 27B INT4 model occupies 24.98~GB of peak
resident memory---37.2\% of the 67.2~GB of RAM---and on Machine B it occupies
20.43~GB, or 59.9\% of 34.1~GB. In both cases the on-board GPU is
out-of-memory: the 16.65~GB weight file alone is $2.6\times$ the 6.4~GB VRAM of
Machine A and $3.9\times$ the 4.3~GB VRAM of Machine B. The same model that no
consumer GPU here can load runs from system RAM with substantial headroom. That
this holds across two CPU microarchitectures, two RAM sizes, two operating
systems, and two GPUs indicates the effect is a property of the memory
hierarchy, not of any single configuration.

\begin{table}[t]
\caption{Per-machine results (INT4). Capacity tier = 27B, edge tier = 2B.}
\label{tab:main}
\centering
\begin{tabular}{@{}llcc@{}}
\toprule
\textbf{Metric} & \textbf{Tier} & \textbf{Machine A} & \textbf{Machine B} \\
\midrule
Peak RSS (GB)        & 27B & 24.98 & 20.43 \\
RAM utilization (\%) & 27B & 37.2  & 59.9 \\
GPU verdict          & 27B & OOM   & OOM \\
\midrule
TTFT (s)             & 27B & 18.4  & 84.7 \\
Throughput (tok/s)   & 27B & 2.36  & 0.62 \\
TTFT (s)             & 2B  & 2.0   & 16.2 \\
Throughput (tok/s)   & 2B  & 19.4  & 2.36 \\
\midrule
CPU power (W, est.)  & --- & 65    & 28 \\
Tokens/joule         & 27B & 0.020 & 0.011 \\
Tokens/joule         & 2B  & 0.212 & 0.052 \\
\bottomrule
\end{tabular}
\end{table}

\subsection{E2: Throughput and latency}
\begin{figure*}[tb]
\centerline{\includegraphics[width=0.95\textwidth]{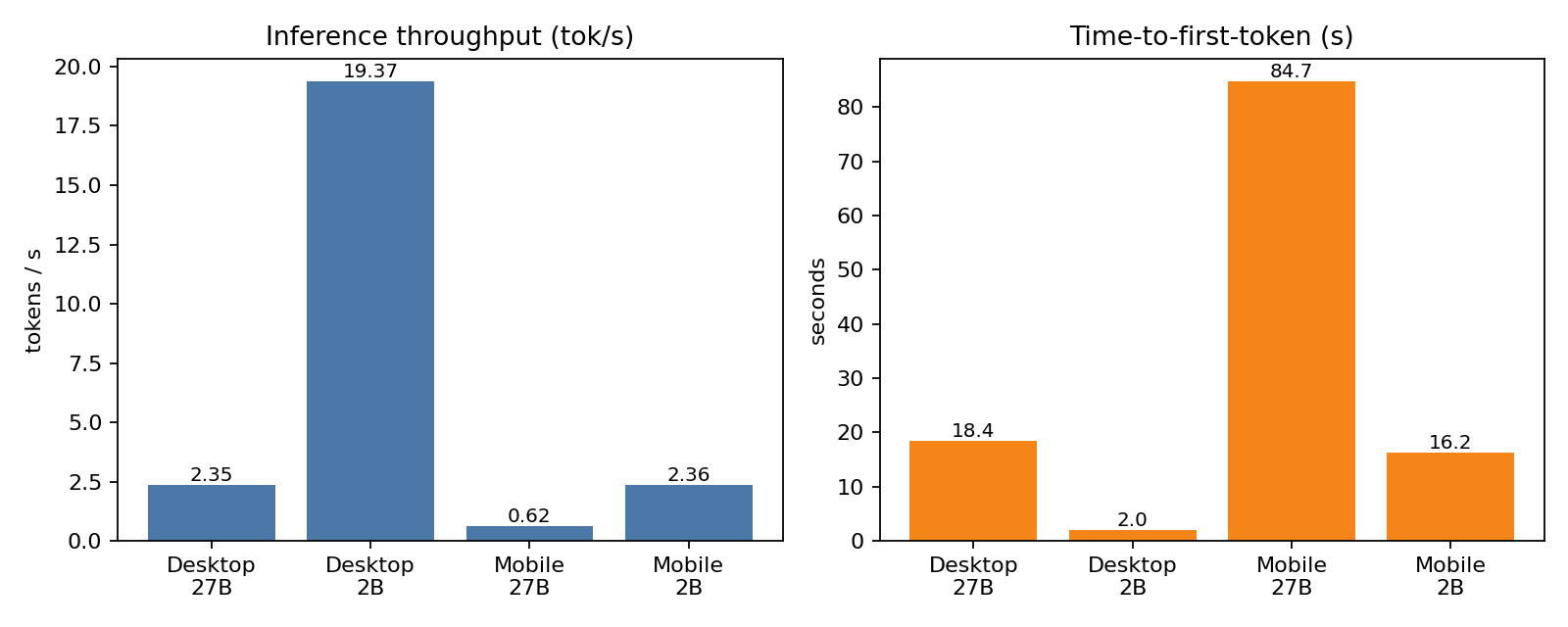}}
\caption{Throughput and time-to-first-token across hosts and model tiers.
Comparisons are meaningful only within a tier; differences reflect CPU class.}
\label{fig:tput}
\end{figure*}

Fig.~\ref{fig:tput} and Table~\ref{tab:main} report inference cost.
Twenty-seven-billion-parameter CPU inference is, as expected,
``overnight-batch'' speed rather than interactive: 2.36~tokens/s on Machine A
and 0.62~tokens/s on Machine B, with time-to-first-token of 18 and 85 seconds.
The 2B edge tier is interactive on the faster machine (19.4~tokens/s).
The gap between the two hosts tracks CPU generation and memory bandwidth (a 2022
desktop part versus a 2018 mobile part), not any accelerator; we present these
as honest CPU-class baselines. The implication is operational: capacity unlocks
\emph{which} models can run locally, and latency dictates \emph{how} they are
used---asynchronous, overnight, or batch summarization rather than real-time
dialogue.

\subsection{E3: Energy}
\begin{figure}[tb]
\centerline{\includegraphics[width=0.86\columnwidth]{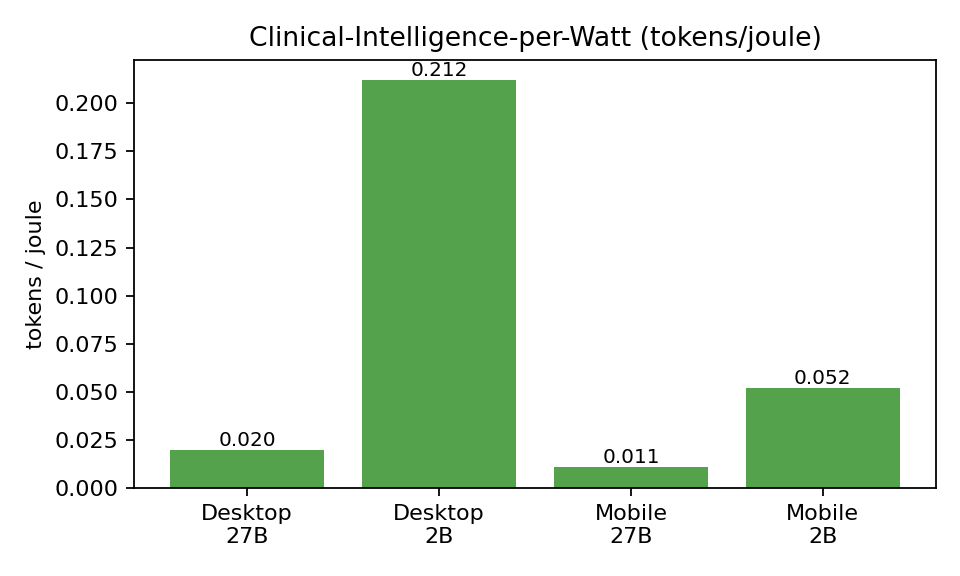}}
\caption{Energy efficiency proxy (tokens per joule). Power is an estimate-grade
TDP figure; values are indicative, not metered.}
\label{fig:ciw}
\end{figure}

Using a labeled TDP estimate (Machine A 65~W, Machine B 28~W), the 27B model
yields 0.020 and 0.011 tokens/joule, and the 2B model 0.212 and
0.052~tokens/joule (Fig.~\ref{fig:ciw}). Even granting the coarse power model,
the figures situate sustained local inference within a ``home-appliance'' power
envelope of tens of watts, in contrast to the 300\,W-class draw of discrete
training GPUs, supporting a continuous, on-premises deployment.

\subsection{E4: Communication efficiency}
In a federated setting, each site shares only its LoRA adapter. For the 27B
model with adapters on the query and value projections, the adapter is 11.3~MB
at rank~8 and 22.6~MB at rank~16, computed from the architecture
(Table~\ref{tab:comms}). Against the $\sim$54~GB of full BF16 weights, this is a
99.98\% and 99.96\% reduction in transmitted volume---decisive for
low-bandwidth clinical edge networks, and a direct consequence of keeping the
heavy base model resident in local RAM rather than moving it.

\begin{table}[t]
\caption{LoRA adapter size vs. full model (27B, q/v projections)}
\label{tab:comms}
\centering
\begin{tabular}{@{}lccc@{}}
\toprule
\textbf{Rank} & \textbf{Adapter (BF16)} & \textbf{Full model} & \textbf{Reduction} \\
\midrule
8  & 11.3 MB & 54.4 GB & 99.98\% \\
16 & 22.6 MB & 54.4 GB & 99.96\% \\
\bottomrule
\end{tabular}
\end{table}

\section{Discussion and Limitations}

The results support a clear deployment thesis: for on-premises clinical LLMs,
system RAM---not GPU VRAM---is the resource that determines which models are
feasible, and a CPU-only path removes the GPU as a gatekeeper while keeping all
data local. The capacity finding (E1) is the robust, machine-independent result;
the cost findings (E2, E3) are honest baselines whose absolute values depend on
CPU class.

Several limitations bound the claims. (i) Throughput is low for the 27B model;
the approach suits batch and asynchronous clinical workflows, not real-time
conversation. (ii) Power is estimated from TDP, not metered; tokens/joule values
are indicative. (iii) The GPU OOM verdict is analytical, established by
weights-versus-VRAM, not by an attempted crash. (iv) We evaluate inference and
analytically size adapters; we do not execute federated fine-tuning end to end.
(v) We use an open Gemma-2 model as a footprint-faithful stand-in for a 27B
medical model and synthetic FHIR data; clinical quality is out of scope and the
artifact is explicitly not a medical device. (vi) Both test CPUs are commodity
parts without matrix-acceleration instruction-set extensions; we make no claim
about, and did not measure, any such accelerator. Newer CPUs with wider vector
or matrix units would be expected to improve E2/E3 but are orthogonal to the
capacity result.

\section{Conclusion}
\label{sec:concl}

We demonstrated, on two independent consumer machines, that a 27B clinical
language model in 4-bit precision loads and serves entirely from system RAM
while the on-board GPU cannot hold it. The result is independent of CPU microarchitecture,
RAM size, operating system, and GPU, reframing the edge node from
compute-limited to capacity-enabled. Combined with LoRA-only federated exchange
(a 99.96\% bandwidth reduction) and a tens-of-watts power envelope, this points
to a private, sustainable, and broadly deployable path for clinical AI on
hardware that already exists in clinics today. Future work proceeds on two
fronts. First, we will port the inference pipeline to the OpenVINO
runtime~\cite{openvino} to exploit its INT4-optimized CPU kernels and to offload
a low-power bedside (mobile) tier onto integrated GPUs and NPUs, converting the
conservative baseline reported here into an optimized, hardware-accelerated
deployment. Second, we will execute end-to-end federated fine-tuning across
sites under the FLAI Protocol's incentive and aggregation layer~\cite{flai}. The
harness, synthetic data generator, and reproduction scripts are released openly
at \url{https://github.com/balkeumlabs/LLM-on-CPU}.

\section*{Acknowledgment}
The authors received no external funding for this work and declare no conflicts
of interest.

\end{document}